\documentstyle[12pt,epsf,axodraw]{article}
\newcommand{\beq}{\begin{equation}}% can be used as {equation} or  {eqnarray}
\def\eeq#1{\label{#1}\end{equation}}
\def\beqa{\begin{eqnarray}}
\def\eeqa#1{\label{#1}\end{eqnarray}}
\def\CR{\nonumber \\ }
\def\leqn#1{(\ref{#1})}

\begin{document} 
%%%%%%%%%%%%%%%%%%%%%%%%%%%%%%%%%%%%%%%%%%%%%%%%%%%%%%
\begin{titlepage}

\begin{flushright}
UFIFT-HEP-09-02 \\ 
FERMILAB-PUB-09-044-T \\
April 10, 2009
\end{flushright}

\vskip 1.5cm

%%%%%%%%%%%%%%%%%%%%%%%%%%%
\begin{center}
{\huge \bf Shedding Light on the Dark Sector with Direct WIMP Production}
\vskip.1cm
\end{center}
\vskip0.2cm

%%%%%%%%%%%%%%%%%%%%%%%%%%%
\begin{center}
{\bf Partha Konar$^1$, Kyoungchul Kong$^2$, Konstantin T. Matchev$^1$, \\
Maxim Perelstein$^3$}
\end{center}

\vskip 8pt

%%%%%%%%%%%%%%%%%%%%%%%%%%%
\begin{center}
{\it $^1$ Physics Department,
        University of Florida,
        Gainesville, FL 32611, USA} \\
\vspace*{0.3cm}
{\it $^2$ Theoretical Physics Department, 
        Fermilab,
        Batavia, IL 60510, USA} \\
\vspace*{0.3cm}
{\it $^3$ Institute for High Energy Phenomenology,\\
Newman Laboratory of Elementary Particle Physics,\\
Cornell University, Ithaca, NY 14853, USA} \\
%\vspace*{0.3cm}
%{\tt Email@email.com}
\end{center}

\vglue 0.3truecm

%%%%%%%%%%%%%%%%%%%%%%%%%%%
\begin{abstract}
\vskip 3pt \noindent A Weakly Interacting Massive Particle (WIMP)
provides an attractive dark matter candidate, and should be within reach
of the next generation of high-energy colliders. We consider the
process of direct WIMP pair-production, accompanied by an initial-state
radiation photon, in electron-positron collisions at the proposed
International Linear Collider (ILC). We present a parametrization of
the differential cross section for this process which conveniently
separates the model-independent information provided by cosmology from
the model-dependent inputs from particle physics. As an application, we
consider two simple models, one supersymmetric, and another
of the ``universal extra dimensions'' (UED) type. The discovery reach of
the ILC and the expected precision of parameter measurements are
studied in each model. 
%For each of the two examples, we perform 
%a shape-discrimination analysis of the photon energy spectrum and show 
%that in a large part of the relevant parameter space it allows 
%to rule out the alternative model interpretation.
In addition, for each of the two examples, we also investigate 
the ability of the ILC to distinguish between the two models
through a shape-discrimination analysis of the 
photon energy spectrum. We show that with sufficient beam 
polarization the alternative model interpretation can be ruled
out in a large part of the relevant parameter space.
\end{abstract}

\end{titlepage}

%%%%%%%%%%%%%%%%%%%%%%%%%%%%%%%%%%%%%%%%%%%%%%%%%%%%%%

%\title{}
%\author{Partha Konar, KC Kong, Konstantin Matchev, Maxim Perelstein}
%\maketitle

\section{Introduction}
\label{sec:intro}
The most convincing and direct evidence for new physics beyond 
the Standard Model (BSM) is the existence of dark matter. 
The evidence for non-baryonic dark matter comes from a variety 
of observations on a wide range of astrophysical scales: 
rotation curves of galaxies, cosmic microwave background, 
gravitational lensing, mass to luminosity ratios from the 
motion of galaxies in clusters of galaxies, power spectrum 
and large scale structure, and so on.
No particle in the Standard Model (SM) can account for the 
observed dark matter. Interestingly, many extensions of 
the SM at the electroweak scale possess good dark matter 
candidates. As the collider experiments begin directly 
probing this energy scale, it is important to understand 
what they can potentially teach us about dark matter.

If one assumes that the dark matter particle was in thermal 
equilibrium with the rest of the cosmic fluid in the early universe, 
and then decoupled once the temperature dropped, 
its present abundance can be predicted. In a generic case, 
the abundance is a function of two continuous parameters:
the dark matter particle mass $M_\chi$ and its {\em total} annihilation 
cross section into SM states at low velocity, $\sigma_{\rm an}$. 
The current cosmological and astrophysical observations provide a measurement 
of the dark matter abundance, and indicate that the preferred values 
of $\sigma_{\rm an}$ are of order 1 pb. Interestingly, this 
also happens to be the typical cross section for processes 
mediated by the SM weak interactions or other weak-scale particles. 
This surprising coincidence provides an additional hint that 
the dark matter particle may indeed by a part of the weak-scale theory. 
The dark matter particles of this class are called Weakly-Interacting 
Massive Particles, or WIMPs.
 
There are two ways in which WIMPs can be produced in high-energy colliders. 
First, they can be directly pair-produced. 
(Since WIMPs are stable due to some new conserved quantum number, 
they cannot be singly produced in the collision of light SM particles.) 
Second, they can appear as decay products of other BSM particles. 
The second channel often has larger rates and smaller backgrounds, 
especially at the LHC where the production of strongly-interacting 
BSM states is enhanced. However, it is important to emphasize that 
this statement is very model-dependent and relies on the presence
of {\em other}, heavier particles in the spectrum, which 
is a strong additional requirement independent of the WIMP dark matter hypothesis. In addition, the interpretation of the data in this channel is challenging. 
For example, the traditional methods for mass determination 
can provide a measurement of the WIMP mass only if long cascade 
decay chains are available and can be identified \cite{Burns:2008va}.
The measurement of the WIMP couplings to the SM states is also a 
very challenging task and even in best-case scenarios one can 
only determine certain combinations of couplings and mixing angles 
\cite{Burns:2008cp}. Moreover, the WIMP may be only one of several
weakly-interacting new particles present in the model, and it may be 
quite challenging to disentangle the features of this sector in 
cascade decays. For these reasons, direct production of WIMPs, 
while more challenging experimentally, may offer unique opportunities 
to study the WIMP in the lab.  

The absence of a robust handle to distinguish events with direct 
WIMP production from the large SM background makes it virtually 
impossible to study this channel at the LHC \cite{us,FST}. 
The relevant signatures (mono-jet plus missing energy 
and single photon plus missing energy) resemble
direct graviton production in ADD models with Large Extra 
Dimensions \cite{Vacavant:2001sd}, which is known to be
among the experimentally most challenging searches at the LHC. 
Both channels are subject to a large irreducible background
from $Z$ + jets/photons, with $Z$ decaying invisibly. In addition, 
there will be a large pure QCD background, where the missing 
energy arises as a detector effect. To make matters worse, 
the signal cross-sections in ADD are typically larger 
than the corresponding cross-sections for direct WIMP production.
As a result of all these factors, the LHC is not expected to
observe WIMPs directly, but only in the cascade decays of other, 
heavier particles. In this paper, therefore,
%In this paper,
we will consider the direct WIMP production at 
the proposed International Linear Collider (ILC), an $e^+e^-$ collider 
with $\sqrt{s}=500$ GeV. Most previous studies have analyzed this
process in the framework of supersymmetry (SUSY), and correspondingly 
refer to it as ``radiative neutralino production'' \cite{Fayet:1982ky,%
Ellis:1982zz,Grassie:1983kq,Kobayashi:1984wu,Ware:1984kq,Bento:1985in,
Chen:1987ux,Kon:1987gi,Chen:1995yu,Ambrosanio:1995it,Choi:1999bs,Baer:2001ia,
Dreiner:2006sb,Dreiner:2007vm,Dreiner:2007qc,Basu:2007ys,Berger:2007ut,Rai:2008ei}.
The related SUSY processes of direct production of sneutrinos or heavier neutralinos
decaying invisibly, have also been considered
\cite{Franke:1994ph,Datta:1994ac,Datta:1996ur,Datta:2002jh}, leading
to the so called ``virtual LSP'' scenarios.
However, the connection between particle physics and cosmology
provided by the WIMP hypothesis is very general, and is not
restricted to supersymmetry. In particular, it may 
be readily applied in other model frameworks which have 
become popular in recent years, e.g. Universal Extra Dimensions (UED)
\cite{Appelquist:2000nn,Servant:2002aq,Cheng:2002ej,Dobrescu:2007ec},
Warped Extra Dimensions \cite{Agashe:2004ci,Agashe:2007jb,Panico:2008bx}
and Little Higgs models with $T$-parity (LHT) \cite{Cheng:2003ju,Birkedal:2006fz}.
Given the multitude of existing models which can accommodate 
a successful WIMP, it is important to address the following two issues:
\begin{enumerate}
\item When predicting the size of any potential signals 
of direct WIMP production at colliders,
how much of the discussion is model-independent, i.e. is relying only
on the {\em generic} WIMP properties required by cosmology.
\item Once a signal of direct\footnote{Model discrimination from 
{\em indirect} WIMP production at the ILC was previously discussed in
\cite{Battaglia:2005zf,Battaglia:2005ma,Bhattacherjee:2005qe,Buckley:2007th}.} 
WIMP production is observed, how can one discriminate among the
various new physics model alternatives, e.g. SUSY, UED, LHT etc. 
\end{enumerate}

The first question was addressed in Ref.~\cite{us}, which
developed the model-independent formalism for predicting the 
rate for direct WIMP production at colliders.
%In a previous publication~\cite{us}, two of us (KM and MP, 
%together with A.~Birkedal) argued that 
Using detailed balancing, the production cross-section
at colliders was related to the leading term $\sigma_{\rm an}$
in the velocity expansion of the WIMP annihilation cross-section
(the procedure will be reviewed below in Section~\ref{sec:direct}).
It was shown that the collider signals can be uniquely predicted in 
terms of only two continuous parameters: the WIMP mass $M_\chi$ and the
WIMP annihilation fraction $\kappa_e$ into the $e^+e^-$ final state
(see the exact definition of $\kappa_e$ in eq.~(\ref{kappadef}) below).
Interestingly, it turned out that the ILC has sensitivity to direct 
WIMP pair-production in an interesting parameter range for $(M_\chi,\kappa_e)$
\cite{us,BL1,ALCPG07,Bernal:2008zk,BL2}, complementing more traditional 
methods for direct or indirect dark matter detection 
\cite{Bernal:2008zk,Birkedal:2005ep}.

While the formalism of Ref.~\cite{us} is attractive due to its model-independence,
it has one significant drawback: in order to use the connection to cosmology,
the WIMP's must be produced at low velocity, i.e. near threshold. On the other hand,
the center-of-mass energy of the ILC may very well be far above threshold.
In order to bring the WIMP system back near threshold, a hard cut on the 
energy $E_\gamma$ of the associated photon was required \cite{us}, which lead to a significant
reduction in the observable signal. In principle, the photon energy cut 
may be relaxed \cite{Bernal:2008zk}, which results in larger signal rates, 
but then the connection to cosmology is lost and the accuracy of the theoretical 
prediction goes awry.

In this paper, we tackle this problem by presenting a slightly modified 
framework to analyze this process. The new framework, outlined in Section~\ref{sec:direct}, 
requires minimal additional inputs from the microscopic model of WIMPs, 
compared to the model-independent formalism of Ref.~\cite{us}. 
The advantage is that it allows to significantly relax the cut on the photon energy, 
extending the ILC reach. We also include the possibility of additional background 
suppression using polarized beams, further improving the reach \cite{Dreiner:2007vm,Dreiner:2007qc}. 
In Section~\ref{sec:mass} we show that once the observation is made, the ILC experiments can provide 
a measurement of the WIMP mass and the cross section of the process 
$e^+e^-\rightarrow \chi\chi$. The latter measurement is especially interesting, 
since if this cross section (at threshold) turns out to be larger than that 
indicated by the WIMP relic abundance, the $\chi$ particle will either be excluded 
as the sole dark matter component, or will require non-standard cosmology
\cite{Chung:2007cn}.

The second goal of this paper is to discuss the second question posed above:
Once a signal of direct WIMP production is seen, how can one determine the 
WIMP properties and correspondingly the particular type of new physics model?
This question has not been adequately addressed in the existing literature, 
since previous works on this channel have only considered one new physics model at a time, 
and did not attempt any model discrimination studies. In our case here,
once our prediction of the observable photon spectrum becomes model-dependent, 
a measurement of this spectrum can be readily used to distinguish between 
different models of WIMPs and their interactions. The results from such an
analysis will be presented in Section~\ref{sec:discrimination}.

\section{Direct WIMP Production at the ILC} 
\label{sec:direct}
 
 In this section, we will define the framework to describe a simple observable 
signature of the direct WIMP production in electron-positron collisions, 
and estimate the expected reach of the ILC in this framework.
 
\subsection{The Signature}

We consider the direct WIMP pair-production process 
\beq
e^+e^-\rightarrow \chi\chi. 
\eeq{process2}
Here, the WIMP $\chi$ is an electrically neutral, color-singlet particle, 
with a mass in the 1 GeV-few hundred GeV range. At this point, 
we do not need to specify any other quantum numbers of $\chi$ (such as its spin), 
or the fundamental theory of which $\chi$ is a part. 
On its own, the process~\leqn{process2} does not leave an observable 
signature in the detector. To get an observable signature, 
we consider the closely related process 
 \beq
e^+e^-\rightarrow \chi\chi\gamma,
\eeq{process3}
 in which the WIMP pair can be observed as missing momentum recoiling 
against the detected photon. The cross section is dominated by the 
photons that are soft or collinear with the beam line: 
\beq
E_\gamma \sin\theta \ll \sqrt{s}\,,
\eeq{scet}
where $E_\gamma$ is the photon energy, $\sqrt{s}$ is the center-of-mass 
energy of the collider and $\theta$ is the angle between the photon and 
the incoming electron. In this regime the differential cross section 
can be factorized according to the Weitszecker-Williams formula:
\beq
\frac{d\sigma_3}{dx \, d\cos\theta} = {\cal F}(x, \theta) \cdot \sigma_2(\hat{s})\,,
\eeq{WW} 
where $\sigma_3 \equiv \sigma(e^+e^-\to\chi\chi\gamma)$, 
$\sigma_2\equiv \sigma(e^+e^-\to\chi\chi)$, $x=2E_\gamma/\sqrt{s}$, 
$\hat{s}=s(1-x)$, and 
\beq
{\cal F} (x, \theta) = \frac{\alpha}{\pi} \frac{1+(1-x)^2}{x}\frac{1}{\sin^2\theta}\,.
\eeq{Fx} 
Of course, very soft or collinear photons remain undetected in a realistic 
detector; however, it was shown in Ref.~\cite{us} that the 
factorization~\leqn{WW} works rather well under realistic experimental conditions. 

In general, the function $\sigma_2(s)$ depends on the microscopic model of 
the WIMP and its interactions with electrons. However, in the region close 
to production threshold, $s-4M_\chi^2\ll s$, this function can be approximated as
\beq
\sigma_2(s) \approx \sigma_0 \left( 1-\frac{4M_\chi^2}{s}\right)^{p+1/2}\,,
\eeq{sigma2th} 
where $p$ is the angular momentum of the lowest partial wave contributing 
to this process. (In particular, $p=0$ for $s$-wave scattering and $p=1$ for $p$-wave.) 
This is the approximation that was used in Ref.~\cite{us}. Its advantages 
are the independence from the underlying microscopic model, 
and the fact that the quantity $\sigma_0$ can be easily related 
to the relic abundance of the $\chi$ particle. The key relation is~\cite{us}
\beq 
\sigma_0 = 2^{2(p-2)}\,(2S_\chi+1)^2\,\kappa_e\,\sigma_{\rm an}\,,
\label{kappadef}
\eeq{s0cosmo}
where $S_\chi$ is the spin of the $\chi$ particle, $\sigma_{\rm an}$ 
is the coefficient of the leading term in the small-velocity expansion 
of the total $\chi$ pair-annihilation cross section, and $\kappa_e$ 
is the fraction of $\chi$ pair-annihilation events at low velocity 
which result in electron-positron pairs. At this point we 
also assume that $\chi=\bar{\chi}$, i.e. that the WIMP $\chi$ 
is identical to its antiparticle. This assumption will be true in our two 
examples discussed explicitly below, but is not really necessary in general --
in the cases where $\chi\ne \bar{\chi}$, there is simply an additional factor of 
$2$ in the right-hand side of (\ref{kappadef}).

The quantity $\sigma_{\rm an}$ 
is directly related to the present relic abundance of the $\chi$ particle. 
Assuming that $\chi$ makes up all of the observed dark matter, 
and using the observed value for the dark matter density 
$\Omega_{\rm dm} h^2 = 0.1143\pm 0.0034$ \cite{Komatsu:2008hk}, 
yields the determination of $\sigma_{\rm an}$ shown in Fig.~1.
%\draftnote{Here we update the "famous" plot using the latest WMAP 
%dark matter density. Or not? Drop this comment if we don't have time to do it.} 
The quantity $\kappa_e$ cannot be obtained from cosmology, and needs 
to be calculated once the model of particle physics is specified. 
Note however that in any model $\kappa_e\leq 1$, so that
eq.~\leqn{s0cosmo} provides a model-independent upper bound
on $\sigma_0$:
\begin{equation}
\sigma_0 \le 2^{2(p-2)}\,(2S_\chi+1)^2\,\sigma_{\rm an}\,,
\label{sigma0bound}
\end{equation}
%the value of $\sigma_0$ given by eq.~\leqn{s0cosmo} 
%provides a model-independent upper bound 
which is valid as long as $\chi$ is the dominant dark matter component.
If the value of $\sigma_0$ measured in collider experiments
turns out to violate the bound (\ref{sigma0bound}), 
there can be only two, equally exciting, explanations:
that in addition to the $\chi$ particle, 
there must exist yet another independent WIMP candidate, or that
there was some sort of non-standard cosmological evolution in the
early universe.

%
%============================================= Figure
\begin{figure}[t]
\centerline{
\epsfxsize= 8.0 cm\epsfysize=7.0cm
                     \epsfbox{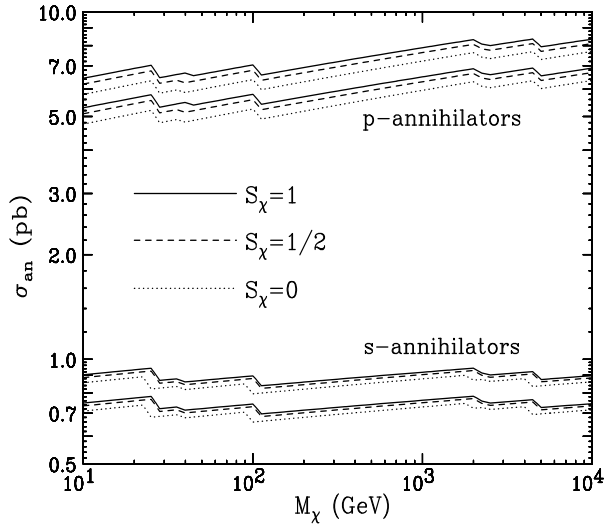}
}
\caption{\sl Values of the quantity $\sigma_{an}$ 
allowed at the $3\sigma$ level as a function of the WIMP mass $M_\chi$,
and for different values of the WIMP spin $S_\chi$. 
The lower (upper) band is for models where
$s$-wave ($p$-wave) annihilation dominates.
We use the constraint $\Omega_{\rm dm} h^2 = 0.1143\pm 0.0034$, which
results from the combination of
data from WMAP-5, Type Ia supernovae and baryon acoustic oscillations
\cite{Komatsu:2008hk}.
}
\label{fig:0}
\end{figure}

The main disadvantage of the approximation~\leqn{sigma2th} is its 
limited range of validity: if one wishes to use it to analyze data, 
one needs to impose a lower bound on the photon energy to restrict 
the analysis to the near-threshold region, thus cutting out most 
of the signal. In this paper, we will pursue a different approach: 
we will calculate $\sigma_2(s)$, for all values of $s$, in two 
benchmark models, and describe the ILC sensitivity for these models. 
As we will see, this gives a substantially better sensitivity than 
the model-independent search of~\cite{us}. To preserve the simple 
connection to relic abundance, we choose to parametrize the cross section as
\beq
\sigma_2(s) = \sigma_0 \,f(s)\,\left( 1-\frac{4M_\chi^2}{s}\right)^{p+1/2}\,,
\eeq{sigma2}       
where $f(s)$ is a model-dependent dimensionless shape function obeying the normalization condition
\beq
f(4M_\chi^2) = 1\,.
\eeq{fnorm}
This form reduces to eq.~\leqn{sigma2th} in the near-threshold region, 
so the coefficient $\sigma_0$ is related to the relic abundance as described above. 

Note that effects such as beamstrahlung \cite{Datta:2005gm} and the possibility of multiple
ISR emission are not included in our approximation scheme. In the future, it 
would be interesting to incorporate these effects in the analysis.

\subsection{Benchmark Models}
%
%============================================= Figure
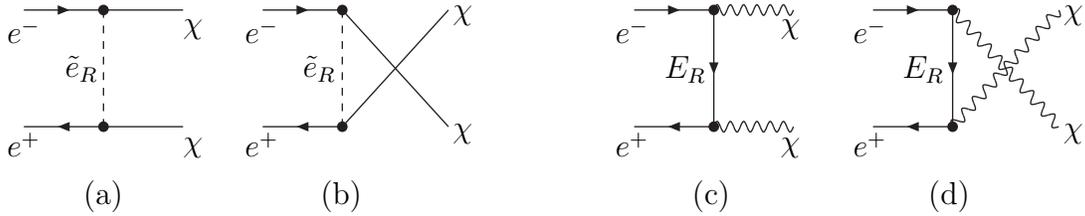
\begin{figure}[t]
\begin{center}
\begin{picture}(403,110)(-4,35)
\ArrowLine(1,109)(31,109)
\ArrowLine(31,65)(1,65)
\DashLine(31,109)(31,65){3}
\Line(31,109)(61,109)
\Line(31,65)(61,65)
\Vertex(31,109){2}
\Vertex(31,65){2}
\Text(1,63)[tc]{$e^+$}
\Text(1,106)[tc]{$e^-$}
\Text(65,61)[tc]{$\chi$}
\Text(65,106)[tc]{$\chi$}
\Text(29,87)[cr]{$\tilde{e}_R$}
\Text(31,45)[tc]{(a)}
\ArrowLine(91,109)(121,109)
\ArrowLine(121,65)(91,65)
\DashLine(121,109)(121,65){3}
\Line(121,109)(161,65)
\Line(121,65)(161,109)
\Vertex(121,109){2}
\Vertex(121,65){2}
\Text(91,63)[tc]{$e^+$}
\Text(91,106)[tc]{$e^-$}
\Text(167,61)[bc]{$\chi$}
\Text(167,106)[bc]{$\chi$}
\Text(119,87)[cr]{$\tilde{e}_R$}
\Text(121,45)[tc]{(b)}
\ArrowLine(231,109)(261,109)
\ArrowLine(261,65)(231,65)
\ArrowLine(261,109)(261,65)
\Photon(261,109)(291,109){2}{6}
\Photon(261,65)(291,65){2}{6}
\Vertex(261,109){2}
\Vertex(261,65){2}
\Text(231,63)[tc]{$e^+$}
\Text(231,106)[tc]{$e^-$}
\Text(291,61)[tc]{$\chi$}
\Text(291,106)[tc]{$\chi$}
\Text(259,87)[cr]{$E_R$}
\Text(261,45)[tc]{(c)}
\ArrowLine(321,109)(351,109)
\ArrowLine(351,65)(321,65)
\ArrowLine(351,109)(351,65)
\Photon(351,109)(391,65){2}{10}
\Photon(351,65)(391,109){2}{10}
\Vertex(351,109){2}
\Vertex(351,65){2}
\Text(321,63)[tc]{$e^+$}
\Text(321,106)[tc]{$e^-$}
\Text(398,61)[bc]{$\chi$}
\Text(398,106)[bc]{$\chi$}
\Text(349,87)[cr]{$E_R$}
\Text(351,45)[tc]{(d)}
\end{picture}
\end{center}
\caption{\sl Feynman diagrams for the WIMP pair-production in the SUSY 
benchmark model (a,b) and the UED benchmark model (c,d).}
\label{fig:FDs}
\end{figure}

Our first benchmark model is supersymmetric~\cite{SUSY_DM}, with a pure Bino LSP
pair-produced in electron-positron collisions via $t$-channel 
exchange of the right-handed selectron $\tilde{e}_R$, see the diagrams in Fig.~\ref{fig:FDs} 
(a,b). This simple toy model approximates the constrained MSSM in the
``bulk'' dark matter region, up to subleading contributions due to 
non-bino admixtures in the neutralino and the heavier $\tilde{e}_L$ exchange. 
In this model, $p=1$, since $s$-wave annihilation at threshold is forbidden 
by CP invariance (up to terms suppressed by $m_e^2/M_\chi^2 \sim 10^{-10}$, 
which we ignore). We obtain
\beqa
\sigma_0 &=& \frac{g^{\prime 4}}{12\pi M_{\tilde{e}_R}^2} 
\frac{z(1+z^2)}{(1+z)^4}\,,\CR
f(y, z) &=& \frac{3y^2 (1+z)^4}{32z^2(1+z^2)(1-y)^{3/2}}
\Bigl[ \frac{4\sqrt{1-y}\left(y(1-z)^2+2z\right)}{y\left(y(1-z)^2+4z \right)}
\CR & & \hskip-1cm +
\left( -1+\frac{1}{z}+\frac{2z}{y(1-z)+2z}\right) \log \frac{y(1-z)+2z(1-\sqrt{1-y})}{y(1-z)+2z(1+\sqrt{1-y})} \Bigr]\,,
\eeqa{xsbino}
where 
\beq
y\equiv \frac{4M_\chi^2}{s},~~z\equiv \frac{M_\chi^2}{M_{\tilde{e}_R}^2}\,.
\eeq{binodefs}
Note that $0 < y, z \leq 1$, and that $\lim_{y\to 1}f(y,z) = 1$, 
as required by eq.~\leqn{fnorm}. This is the cross section for unpolarized beams; since only 
right-handed electrons couple to $\tilde{e}_R$, the corresponding cross section for 
polarized beams can be found by simply multiplying by an overall factor of
\begin{equation}
\left(1+P_{e^-}\right) \left(1+P_{e^+}\right)\ ,
\end{equation}
where 
\beqa
P_{e^-} = \frac{n(e^-_R)-n(e^-_L)}{n(e^-_R)+n(e^-_L)},~~
P_{e^+} = \frac{n(e^+_L)-n(e^+_R)}{n(e^+_L)+n(e^+_R)}. 
\eeqa{epol}
are the electron and positron beam polarizations, respectively.

The second benchmark model is a spin-1 WIMP, pair-produced in 
electron-positron collisions via $t$-channel exchange of the 
spin-1/2 ``heavy electron'' $E_R$, see the diagrams in Fig.~\ref{fig:FDs}(c,d). 
The heavy electron is a Dirac fermion, and is assumed to be 
the partner of the right-handed electron $e_R$; this assumption
fixes the helicity structure of the vertices in Fig.~\ref{fig:FDs}(c,d).
This toy model reproduces the leading contribution to this process in the 
minimal UED model~\cite{Servant:2002aq,Kong:2005hn,Burnell:2005hm}.\footnote{The 
unpolarized cross section is identical in the 
minimal Littlest Higgs model with T parity~\cite{LHT_DM,Birkedal:2006fz}, 
but the helicity of 
the vertices is the opposite, since only the left-handed electron has a 
partner in that model. Note also that the value of $\kappa_e$ in the LHT model
is small, of order $10^{-4}$, due to a group theory factor suppression of the
vertices in Fig.~\ref{fig:FDs}(c,d).} 
In this case, $s$ annihilation is allowed, and $p=0$. We obtain
\beqa
\sigma_0 &=& \frac{g^{\prime 4} Y_{E_R}^4}{128\pi M_{E_R}^2} 
\frac{z}{(1+z)^2},\CR
f(y, z) &=& \frac{y(1+z)^2}{128z^5\sqrt{1-y}\Bigl(y(1-z)+2z\Bigr)\Bigl(y(1-z)^2+4z\Bigr)} \CR 
& & \hskip-1cm \Bigl[4z\sqrt{1-y} \Bigl(y(1-z)+2z\Bigr) \Bigl(y(1-z)^2(8z^2+3)+4z(8z^2-4z+3)\Bigr) \CR  
& & \hskip-1cm -\, \Bigl(y^2(1-z)^2(8z^2+3)+4yz(8z^2-6z+3)+8z^2(4z^2+1)\Bigr)\CR
& & \hskip-0.6cm \Bigl(y(1-z)^2+4z\Bigr) \log \frac{y(1-z)+2z(1-\sqrt{1-y})}{y(1-z)+2z(1+\sqrt{1-y})} \Bigr]\,,
\eeqa{xsbprime}
where
\beq
y\equiv \frac{4M_\chi^2}{s},~~z\equiv \frac{M_\chi^2}{M_{E_R}^2}\,.
\eeq{bprimedefs}
Note that we will continue to use the ``generic'' WIMP notation, $\chi$, in 
the UED case; the $\chi$ in this case is actually the spin-1 massive partner 
of the hypercharge gauge boson, commonly known as $B^1$ \cite{Cheng:2002iz}.

Of course, it is straightforward to apply our formalism to other models, 
and we expect qualitatively similar results, as long as WIMPs do have 
an unsuppressed coupling to electrons. In all such cases the starting point
of the analysis would still be the parametrization (\ref{sigma2}), from where
the current discussion can be generalized in two different ways. First, 
one may consider a different region of parameter space of the same model 
(SUSY or UED), where there can be additional relevant diagrams beyond those
shown in Fig.~\ref{fig:FDs}. For example, the addition of a $t$-channel 
$\tilde e_L$ exchange would simply modify the functional form of (\ref{xsbino})
and introduce an extra parameter (the mass of $\tilde e_L$). Alternatively, 
one may consider a different model altogether, by deriving the corresponding
function $f(s)$ relevant for that case. From that point on, the analysis 
would still follow the steps outlined below for our two model examples (SUSY and UED).

\subsection{ILC Reach}
\label{sec:reach}

The above formalism provides a prediction for the differential rate of $\gamma+$missing 
energy events at an $e^+e^-$ collider in terms of two continuous parameters, 
$M_\chi$ and $\sigma_0$, once a specific model is chosen to fix the function $f(s)$. 
Experiments at an $e^+e^-$ collider can search for this signature as 
an excess over the SM background. We analyzed the reach of a 500 GeV ILC, 
assuming the following kinematic acceptance cuts:
\beq
\sin\theta \geq 0.1,~~~p_{T,\gamma}\equiv E_\gamma \sin\theta > 3~{\rm GeV.}
\eeq{cuts}
The $p_{T,\gamma}$ cut is imposed to reject backgrounds such as $e^+e^-\to
e^+e^-\gamma$ in the region where the $e^+$ and $e^-$ are too forward to be
detected. This cut corresponds to the improved BeamCal acceptance of 6.67 mrad
for 500 GeV linear collider~\cite{Bambade:2006qc}. With these cuts, the SM background is 
strongly dominated by the process $e^+e^-\to\nu\bar{\nu}\gamma$, and we ignore 
all other backgrounds in this study. We 
simulated this background using {\tt CalcHEP} \cite{Pukhov:2004ca} and 
{\tt CompHEP} \cite{Pukhov:1999gg}. To quantify the 
reach, we compare the differential cross sections $d\sigma/dE_\gamma$ for
signal+BG and pure BG. To account for finite detector resolution, we smear 
the energy of the photon spectrum for both signal and background according to 
$\delta E/E=14.4\%/\sqrt{E}_{\rm GeV} \oplus 0.5\%$~\cite{BL1}. We bin the
events in 5 GeV bins in the photon energy, and compute the event numbers in 
each bin, $\{N^i_{\rm sig+BG}\}$ and $\{N^i_{\rm BG}\}$, for a given 
integrated luminosity ${\cal L}_{\rm int}$. We then use the standard $\chi^2$ 
technique to quantify the probability of measuring the event numbers predicted 
by the BG-only model if the true model is signal+BG. We define
\beq
\chi^2 \,=\, \sum_{i=1}^{N_{\rm bin}} 
\frac{(N^i_{\rm sig+BG}-N^i_{\rm BG})^2}{N^i_{\rm BG} +(\delta_{\rm sys} 
N^i_{\rm BG})^2}\,,
\eeq{chi2}
where $\delta_{\rm sys}$ is the fractional systematic uncertainty in the 
background prediction. Note that we conservatively assume that this systematic
error is uncorrelated between bins. 

%
%============================================= Figure
\begin{figure}[t]
\centerline{
\epsfxsize= 8.0 cm\epsfysize=7.0cm
                     \epsfbox{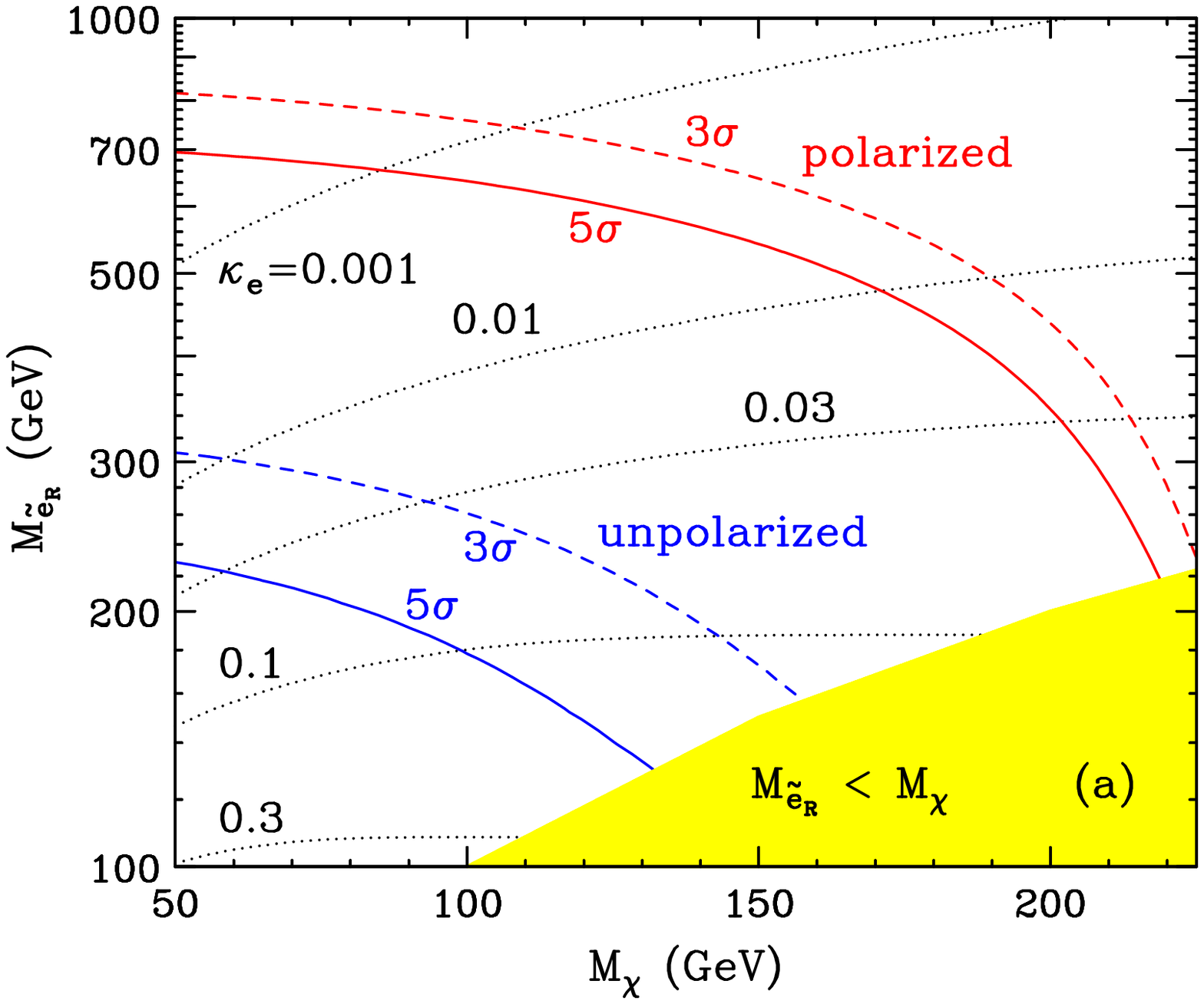}
        \hspace*{.2cm}
\epsfxsize=8.0 cm\epsfysize=7.0cm
                     \epsfbox{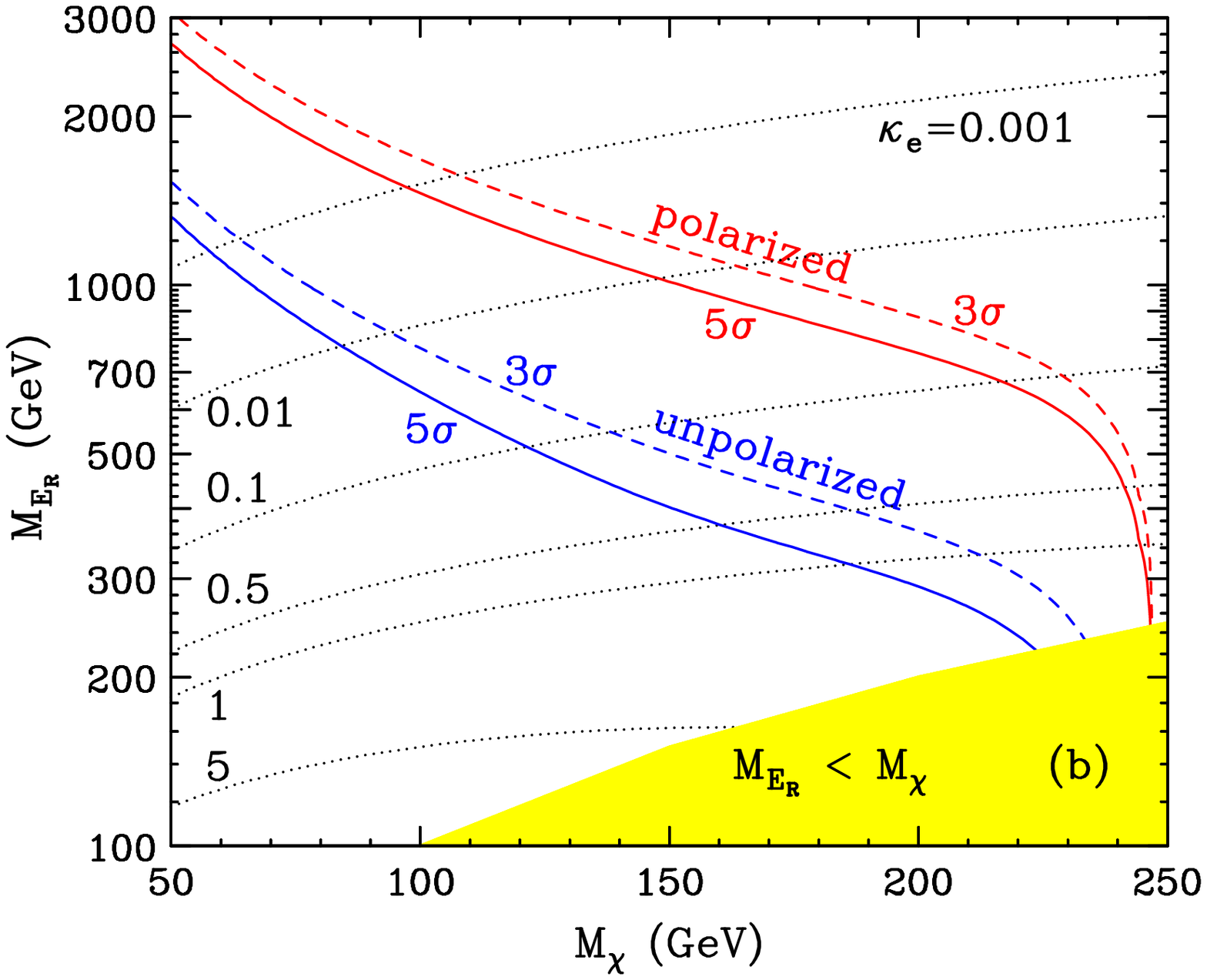}
}
\caption{\sl 
The reach of a 500 {\rm GeV} electron-positron collider with an
integrated luminosity of ${\cal L}_{\rm int} = 500\ {\rm fb}^{-1}$ for discovery 
of our two benchmark models: (a) SUSY and (b) UED. Blue (red) lines correspond to 
unpolarized (polarized with $P_{e^-}=0.8$, $P_{e^+}=0.6$) beams. 
Solid (dashed) lines indicate $5\sigma$ ($3\sigma$) significance. 
Also shown are $\kappa_e$ contours in the (a) $(M_\chi,M_{\tilde e_R})$ 
or (b) $(M_\chi,M_{E_R})$ parameter plane. 
}
\label{fig:1}
\end{figure}

The reach of the search in the benchmark models defined above is presented 
in Fig.~\ref{fig:1}. We assumed an integrated luminosity of 
${\cal L}_{\rm int} = 500\ {\rm fb}^{-1}$, and a
systematic error on the background prediction $\delta_{\rm sys}=0.3$\%. 
The 3$\sigma$
evidence and 5$\sigma$ discovery contours correspond to $\Delta\chi^2=9$ and
$25$, respectively. The contours labeled ``polarized'' correspond to 
electron and positron beam polarizations
of $P_{e^-}=0.8$ and $P_{e^+}=0.6$, correspondingly. 
Note that these polarizations strongly suppress 
the dominant contribution to the background from $t$-channel $W$ exchange, 
which only appears for left-handed electrons, while at the same time 
enhancing the signal in both of our benchmark models, where only the 
right-handed electrons couple to WIMPs. 
The reach is presented in terms of the WIMP mass and the mass of the 
$t$-channel particle which dominates the production process. 
One should keep in mind that the $t$-channel particle can also be directly
produced at the ILC, provided its mass is within the kinematic reach 
of the collider. In this sense, one should note that the 5$\sigma$
reach with unpolarized beams in the SUSY case of Fig.~\ref{fig:1}(a)
only includes regions where the right-handed slepton $\tilde e_R$ 
would be directly produced as well. Consequently, in order to access 
the remaining parameter space where the radiative neutralino production 
is {\em the only} available signal, one must rely on beam polarization.
In Fig.~\ref{fig:1} we also show
contours of constant $\kappa_e$, defined in eq.~\leqn{s0cosmo}. For low
WIMP masses, the reach of the ILC is impressive: with polarized beams, the
WIMPs below 100 GeV can be discovered even if the $e^+e^-$ channel only
contributes 0.1\% to the total WIMP pair-annihilation cross section in the early
universe. For larger WIMP masses, the sensitivity decreases due to the 
kinematic suppression in WIMP production. This suppression can of course be 
compensated by a higher center-of-mass energy. 

\section{Parameter Measurement}
\label{sec:mass}

Once the presence of non-SM contribution in the $\gamma+$missing energy 
channel is established, one can attempt to fit this contribution with BSM model predictions.
In each model, the prediction depends on a number of unknown parameters; 
the fit to data can be used to constrain these parameters. In this section, we
will discuss the accuracy of this procedure, using our benchmark models as 
examples. (Of course, it is
possible that in a given model a good fit cannot be obtained for {\it any}
sensible parameter values; in this case, the model would be ruled out as the 
interpretation of the data. We will discuss examples of such model
discrimination in the next section.) 

\begin{figure}[t]
\centerline{
\epsfxsize= 8.0 cm\epsfysize=7.0cm
                     \epsfbox{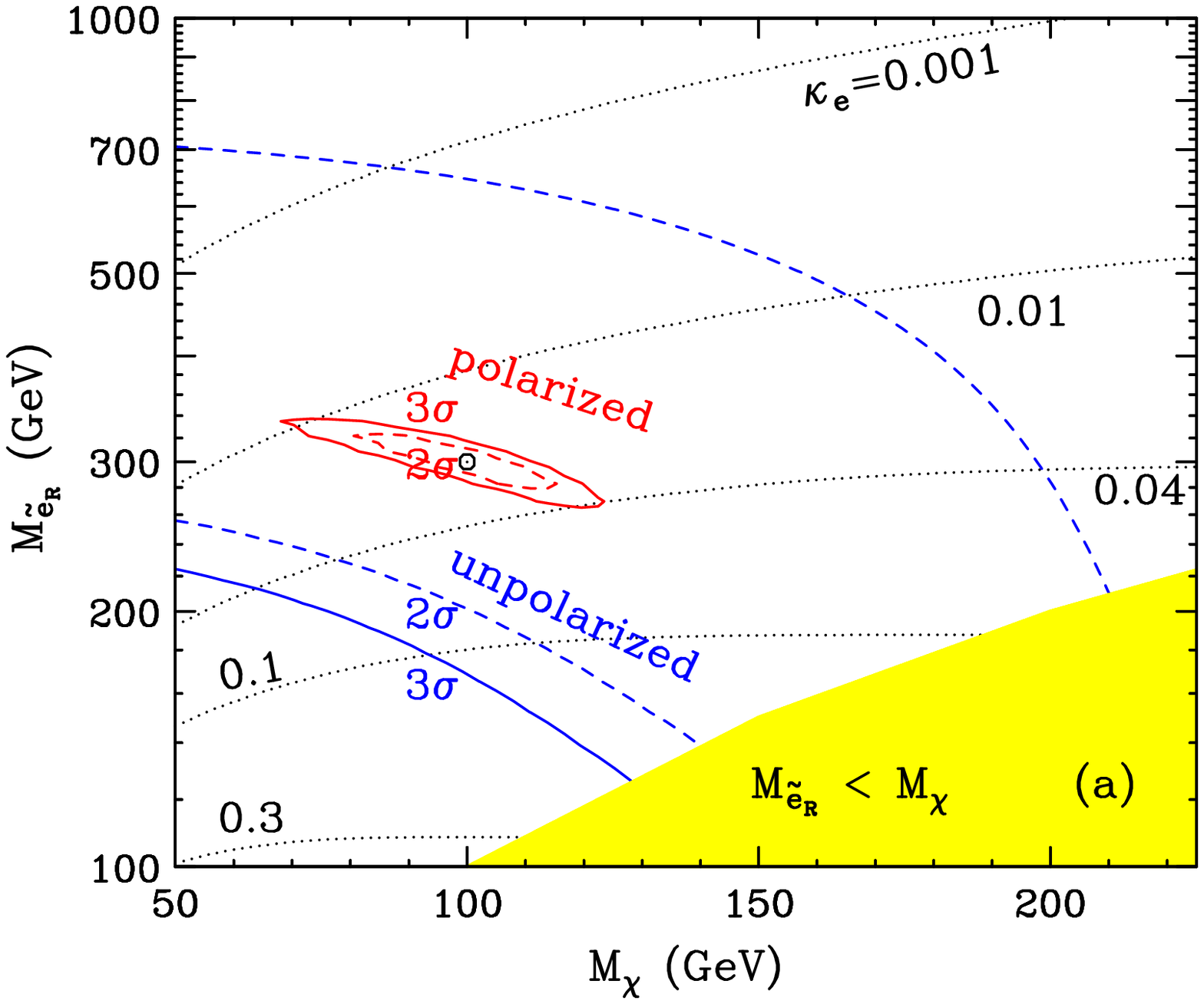}
        \hspace*{.2cm}
\epsfxsize=8.0 cm\epsfysize=7.0cm
                     \epsfbox{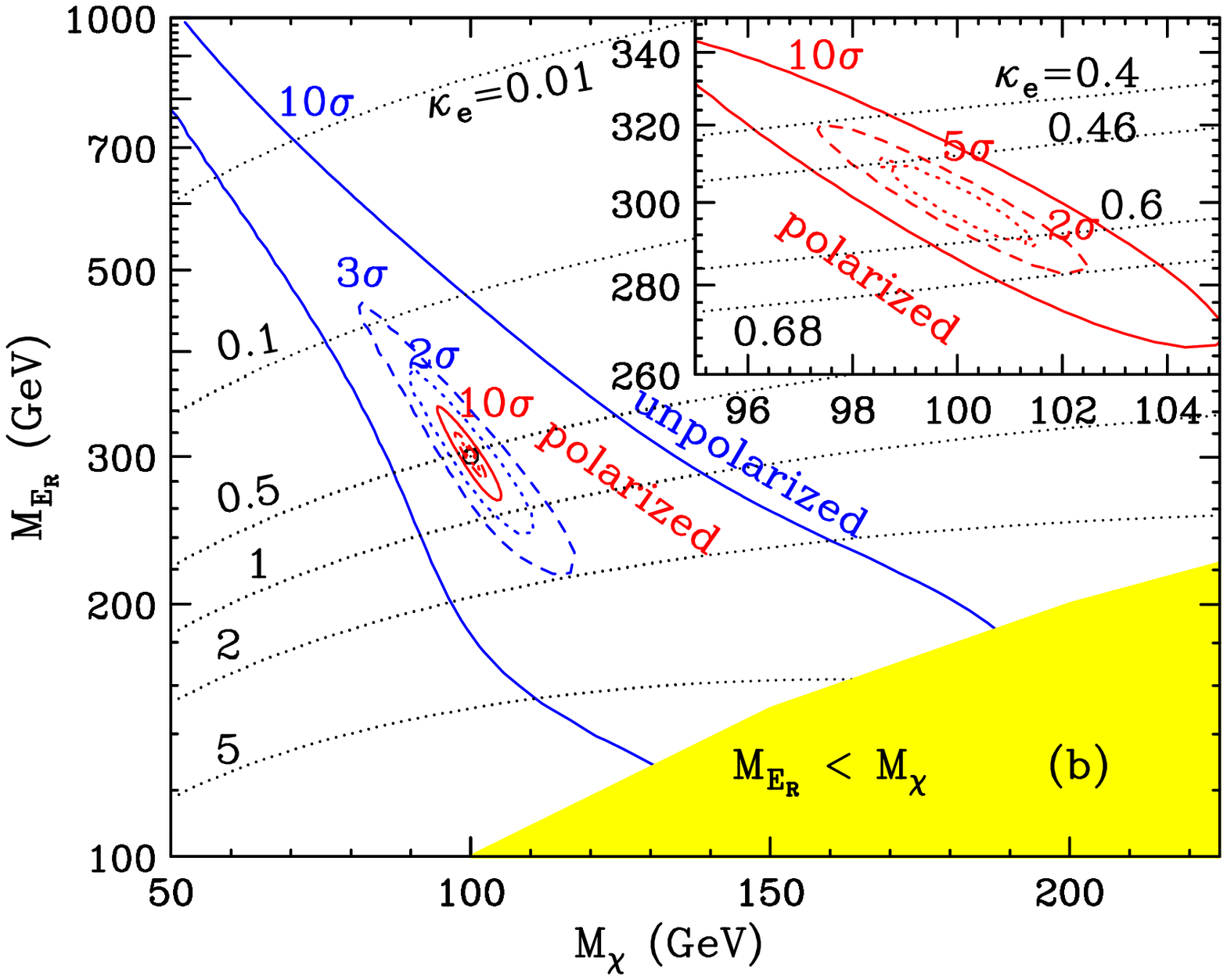}
}
\caption{\sl Expected precision of the model parameter measurement 
in the two benchmark scenarios: (a) SUSY and (b) UED.
The notation and labelling are the same as in Fig.~\ref{fig:1}.}
\label{fig:2}
\end{figure}

In the first example, we assumed that the data is given by the prediction
of the SUSY benchmark model (plus SM background), with 
\beq
M_\chi = 100~{\rm GeV},~~~M_{\tilde{e}_R} = 300~{\rm GeV}.
\eeq{SUSYmasses}
For reference, these values correspond to $\kappa_e=0.023$. 
We then performed a scan of the model parameter space,
varying $M_\chi$ between 50 and 250 GeV, and $M_{\tilde{e}_R}$ between
100 and 1000 GeV. (We imposed $M_{\tilde{e}_R}>M_\chi$.) For each point in 
the scan, we computed the probability of 
observing the data given the model parameters at that point. To quantify this
probability, we used the same $\chi^2$ technique as in the reach analysis
above. The results are presented in Fig.~\ref{fig:2}(a). With unpolarized
beams, only a very crude mass determination is possible: for example, any
WIMP mass up to 220 GeV is allowed at a 2-sigma level. With 
polarized beams, however, a rather impressive accuracy of about $\pm 20$ GeV in both
$M_\chi$ and $M_{\tilde{e}_R}$, at a 2-sigma level, can be achieved. 

%
%
%============================================= Figure

In Fig.~\ref{fig:2}(a) we also show contours of the quantity $\kappa_e$,
which is uniquely fixed by eq.~(\ref{kappadef}), once the values of 
$M_\chi$ and $M_{\tilde{e}_R}$ are specified. Therefore, a 
measurement of the parameters $(M_\chi,M_{\tilde{e}_R})$ can 
be immediately reinterpreted as a $\kappa_e$ measurement.
In particular, from Fig.~\ref{fig:2}(a) we see that with unpolarized beams,
the ILC will only set an upper bound on $\kappa_e$ of about $0.22$ at the
$3\sigma$ level, while using the polarized beam option, 
the $\kappa_e$ determination is much more precise:
$0.01 \le \kappa_e\le 0.04$ at $3\sigma$.

In the second example, we repeated the same exercise for the UED benchmark 
model. The results, presented in Fig.~\ref{fig:2}(b), are
qualitatively similar. The main difference in this case is quantitatively 
better sensitivity, mainly due to a higher signal cross section for the 
chosen model point ($M_\chi = 100$~GeV, $M_{E_R} = 300$ GeV, which in this 
case corresponds to $\kappa_e=0.53$). As a result, a fairly
accurate measurement of the masses is possible even with unpolarized beams:
the accuracy on the WIMP mass is of order $\pm 10$ GeV ($\pm 20$ GeV) at the 2-sigma (3-sigma) level. 
With polarized beams, a WIMP mass determination at the level of $\pm 2$ GeV 
becomes possible. The corresponding 2-sigma determination of $\kappa_e$ is 
now $0.2 \le \kappa_e\le 1.3$ ($0.46 \le\kappa_e \le 0.6$) with 
unpolarized (polarized) beams.

Note that the capability of the ILC to measure the WIMP mass in this
process has also been recently analyzed by Bartels and List~\cite{BL2}. 
Their results for the spin-1 case (the only one analyzed in their study) are 
qualitatively in agreement with ours. Precise agreement is not expected,
since the two studies used a different choice of the benchmark point, as well 
as a slightly different statistical analysis procedure - for example, 
Ref.~\cite{BL2} fits the WIMP mass $M_\chi$ assuming that $\kappa_e$ 
is already known, whereas we perform a 2-parameter fit. Our results are
also in qualitative agreement with those of Refs.~\cite{ALCPG07,Bernal:2008zk},
which used the model-independent framework of \cite{us} to perform a
2-parameter fit for $(M_\chi,\kappa_e)$. 

\section{Model Discrimination}
\label{sec:discrimination}

Since the shape of the photon spectrum in the events with WIMP production 
is model-dependent (as characterized by the function $f(s)$), measuring this
shape can provide interesting information about the underlying model. 
While not an unambiguous spin determination, this measurement can 
nevertheless distinguish scenarios in which WIMPs have different spins, as
long as the WIMP interactions are specified in each model. As always, it is
important to keep in mind that each of the models has free parameters, 
and a model can be ``ruled out'' only if the point in its {\em full} parameter space  
providing the {\it best fit} to data still does not provide an acceptable
fit. (This point has been recently emphasized in Refs.~\cite{Csaki:2007xm,Burns:2008cp,Hallenbeck:2008hf}.)

%
%
%============================================= Figure
\begin{figure}[tb]
\centerline{
\epsfxsize= 8.0 cm\epsfysize=7.0cm
                     \epsfbox{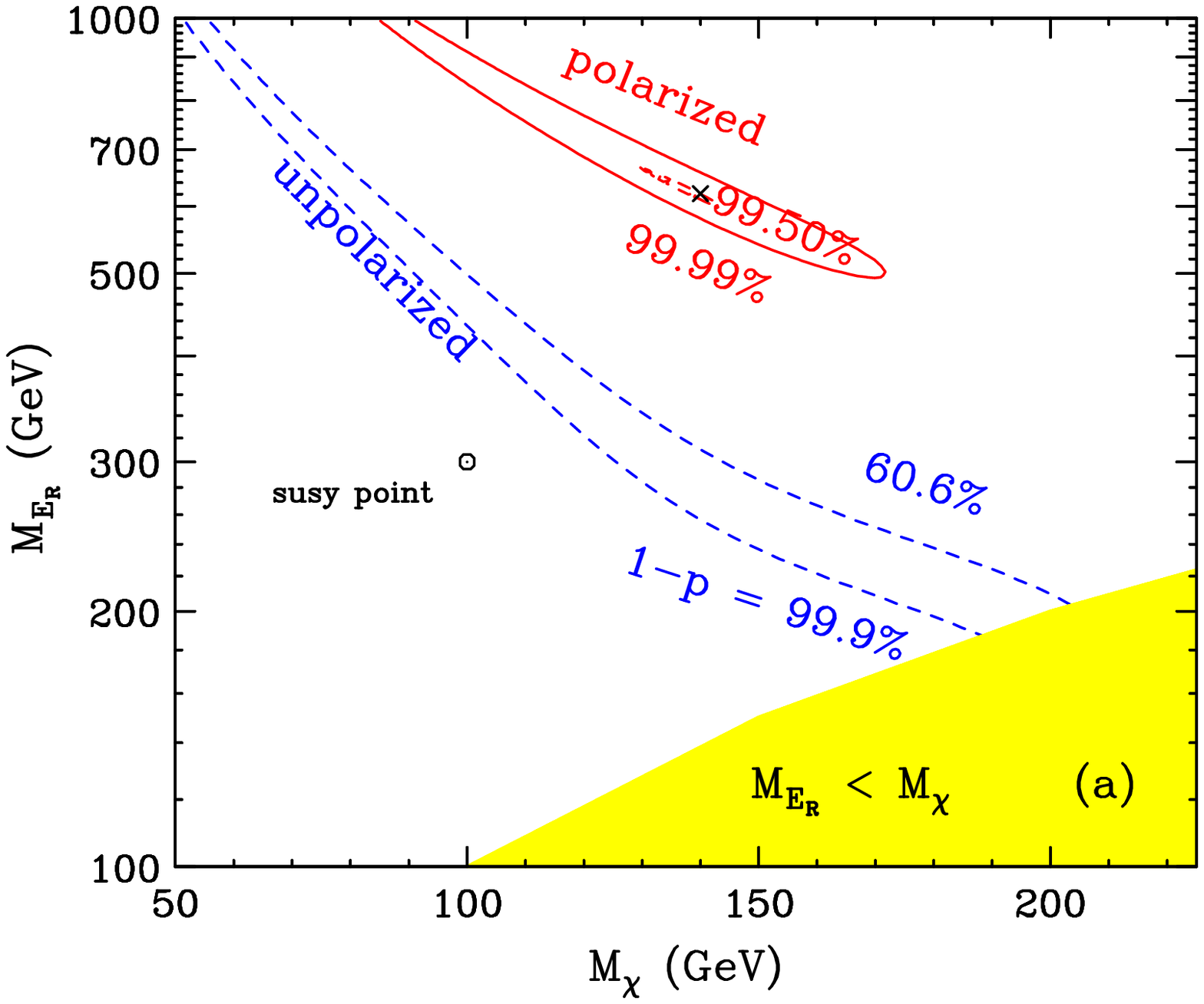}
        \hspace*{.2cm}
\epsfxsize=8.0 cm\epsfysize=7.0cm
                     \epsfbox{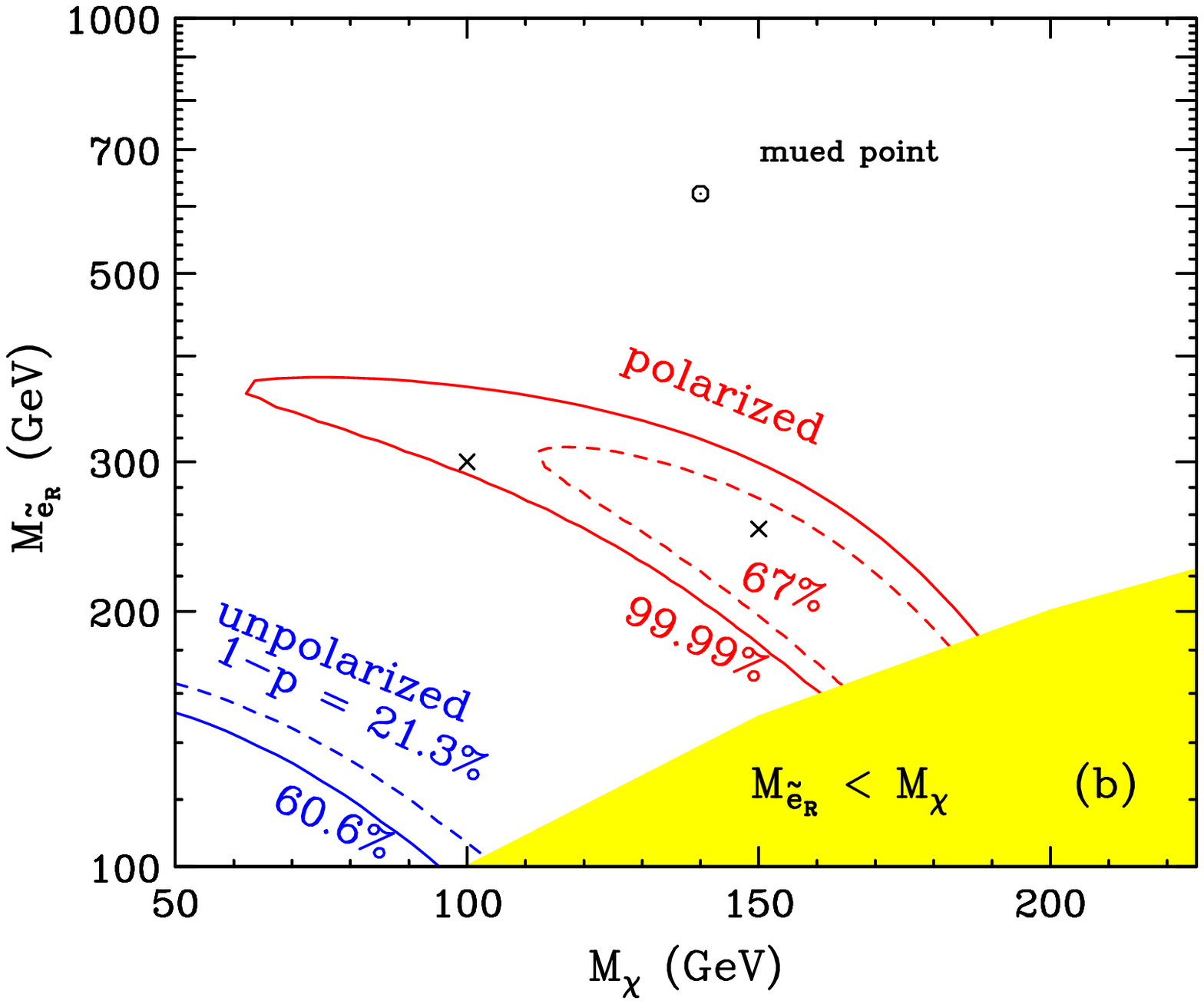}
}
\caption{\sl Plots illustrating the ability of the ILC to discriminate between 
the two benchmark scenarios using the radiative WIMP production process.}
\label{fig:3}
\end{figure}

\begin{figure}[tb]
\centerline{
\epsfxsize= 8.0 cm\epsfysize=7.0cm
                     \epsfbox{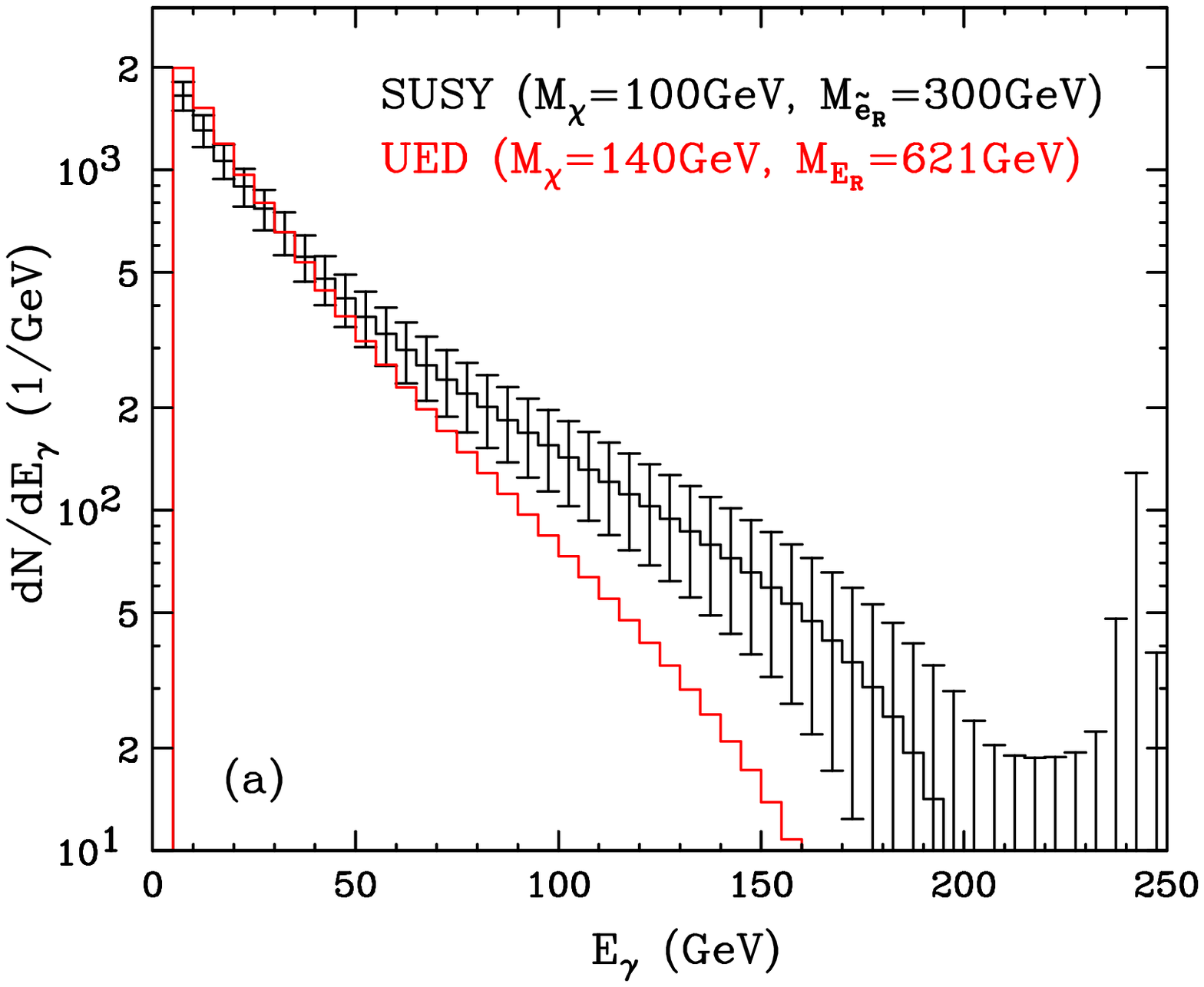}
        \hspace*{.2cm}
\epsfxsize=8.0 cm\epsfysize=7.0cm
                     \epsfbox{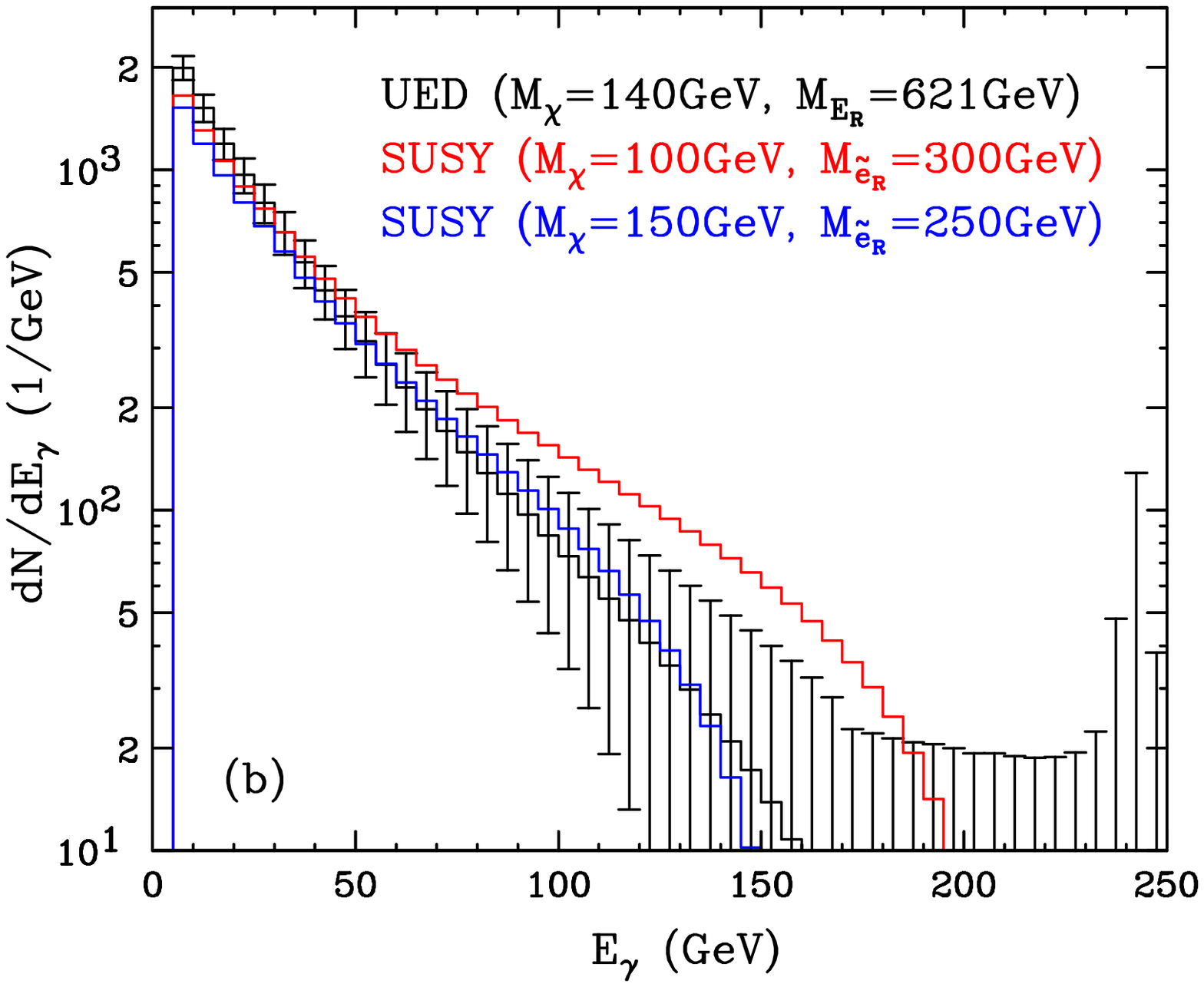}
}
\caption{\sl Signal photon spectra in the two benchmark models. (a) Black
histogram: SUSY model with the parameters in eq.~\leqn{SUSYmasses}; red   
histogram: UED model with parameters giving the best fit to the SUSY
model. (b) Black histogram: UED model with the parameters in 
eq.~\leqn{ued_masses}; blue and red histograms: SUSY models with parameters 
as indicated by the two crosses in Fig.~\ref{fig:3}(a). In both cases, 
polarized electron and positron beams are assumed. The error bars on these
``signal'' histograms are derived from the full signal+BG distributions,
with the statistical (${\cal L}_{\rm int}=
500$ fb$^{-1}$) and systematic (0.3\%) errors, added in quadrature.} 
\label{fig:4}
\end{figure}

To illustrate the power of the ILC $\gamma$+missing energy measurements to
distinguish between the two benchmark models, we performed two sample studies.
In the first one, we assume that the data is given by the prediction
of the SUSY benchmark model (plus SM background), with the same parameters as
in eq.~\leqn{SUSYmasses}. We then attempt to fit the data with the predictions
of the UED model, varying the parameters $M_\chi$ and $M_{E_R}$. For each
parameter point, we determine the $\chi^2$ value, and then compute the 
conditional probability $p$ that the UED model with these parameters is true, 
given the data. The results are shown in Fig.~\ref{fig:3}(a).  
With unpolarized beams, the model-discriminating power of this measurement
is rather poor: any UED model with parameters above the upper blue 
(dashed) contour could be true with probability above 40\%. The UED model
needs to be in the high-mass region to match the observed overall event rate,
but since this rate is rather low and the background is high, there is not 
much sensitivity in this case. The model-discriminating power is greatly
enhanced by beam polarization, which strongly suppresses the background and 
(in both of our benchmark models) enhances the signal. With polarized 
beams, the UED interpretation can be safely ruled out:
even the best-fit point in the UED space (which happens to be at
$M_\chi=140$ GeV, $M_{E_R}=621$ GeV) has only a 0.5\% probability of 
describing the data. The photon spectra for the data (with polarized beams) 
and the best-fit UED 
point are shown in Fig.~\ref{fig:4}(a); it is clear from this figure that 
the quality of the UED fit to data is rather poor. One can conclude that the
UED benchmark model as a whole is ruled out at a 99.5\% CL by this data.  

%
%
%============================================= Figure

We repeated this analysis assuming that the true model is UED, with
\beq
M_\chi = 140~{\rm GeV},~~M_{E_R}= 621~{\rm GeV},
\eeq{ued_masses}
and attempting to fit it with the SUSY predictions. The results are  
shown in Fig.~\ref{fig:3}(b). In this case, the 
model-discriminating power of the measurement is significantly weaker, 
primarily due to the low signal cross section. In the unpolarized case, 
most of the SUSY parameter space is allowed; only the points with   
low $M_\chi$ and $M_{\tilde{e}_R}$ are constrained. The situation is
improved with the polarized beams, where most of the SUSY parameters 
space is ruled out at a confidence level of 99.9\% or better. However,
a significant part of the parameter space survives. This is clear from the
photon spectra shown in Fig.~\ref{fig:4}: the ``data'' spectrum is well fit, 
within the error bars, by the predictions of sample SUSY points. 
The best-fit SUSY point, at $M_\chi=170$ GeV, $M_{E_R}=200$ GeV, 
gives a photon spectrum which is essentially identical to the input
UED ``data''. Thus, convincing discrimination
between the two models appears impossible in this case.  
However, this rather pessimistic conclusion is largely due 
to our choice of a relatively heavy UED study point. 
When we repeat the same analysis for the UED study point
used in Fig.~\ref{fig:2}(b), 
%with the same mass spectrum (\ref{SUSYmasses}) as in our SUSY example,
namely $M_\chi = 100$ GeV and $M_{E_R}= 300$ GeV,
we find that the SUSY interpretation is now completely ruled out.

The two models selected for our study differed in the behavior of the cross sections in the non-relativistic limit (s-wave for UED vs. p-wave for SUSY). This leads to a large difference in the normalization factors $\sigma_0$ in the two models, see Eq.~\leqn{sigma0bound}. However, by varying the masses, the overall rates of the two models can be made to agree, and it is the photon spectrum shape difference that is crucial for breaking that degeneracy. If instead we considered two models that both had, for example, s-wave annihilation, then the normalization $\sigma_0$ would be the same, and one would need to rely exclusively on the spectrum shape differences. As long as the spectra predicted by the two models are sufficiently different, model discrimination should be possible. Thus, the proposed technique should be useful in discriminating between models with the same
threshold behavior of the WIMP annihilation cross section. 

\section{Conclusions}
\label{sec:conclusions}

The ILC may offer a unique window into the dark matter world, 
by providing an opportunity to {\em directly} produce dark matter 
particles and study their properties. 
In this paper we revisited what is perhaps the 
most conservative and least model-dependent collider signature 
of dark matter: direct pair production of WIMPs, tagged with a
photon from initial state radiation \cite{us}. Our main results 
are as follows:
\begin{itemize}
\item We extended the formalism of Ref.~\cite{us} into the 
region away from the endpoint of the photon energy spectrum.
The advantage of this new, although somewhat more model-dependent, 
approach is that it utilizes all of the available information
from the single photon data, and not just near the endpoint.
As a result, the discovery reach of the ILC is significantly 
improved, since one is going beyond a simple counting experiment, 
and using the measured shape of the spectrum as well.
\item In Section~\ref{sec:reach} we estimated the ILC discovery 
reach for two representative models: SUSY and ``UED''-like.
In both cases we find that radiative WIMP production can be 
observable, as long as the WIMP mass is within the kinematic reach
of the ILC and the WIMPs have non-negligible couplings to electrons.
It is worth emphasizing the large difference in the reach with 
and without beam polarization -- see Fig.~\ref{fig:1}.
\item As demonstrated in Sec.~\ref{sec:mass}, the ILC 
can determine the masses of both the WIMP and the associated 
$t$-channel particle (in our two examples, a slepton or a KK-lepton).
This measurement is particularly impressive if beam polarization 
is available -- see Fig.~\ref{fig:2}.
\item Having observed a signal of radiative WIMP production,
the ILC may also attempt a discrimination among different model 
explanations. The outcome of this analysis in general depends on
the specific region of parameter space where the data happens to lie,
but as a rule, beam polarization significantly improves the chances 
for success -- see Fig.~\ref{fig:3}.
\end{itemize}

\vskip0.5cm

\noindent{\large \bf Acknowledgments} 

We thank Hye-Sung Lee for collaboration in the early stage of this project.
KM and MP would like to thank the Kavli Institute for Theoretical Physics 
(KITP) in Santa Barbara, where parts of this work were completed. 
PK and KM are supported in part by a US Department of Energy 
grant DE-FG02-97ER41029. MP is supported by the NSF grant PHY-0355005. 
Fermilab is operated by Fermi Research Alliance, LLC under
Contract No. DE-AC02-07CH11359 with the U.S. Department of Energy.

\end{document}